# Modulation-Doping a Correlated Electron Insulator

Debasish Mondal[1], Smruti Rekha Mahapatra[1], Abigail M Derrico[2], Rajeev Kumar Rai[3], Jay R Paudel[2], Christoph Schlueter[4], Andrei Gloskovskii[4], Rajdeep Banerjee[1], Frank M F DeGroot[5], Dipankar D Sarma[1], Awadhesh Narayan[1], Pavan Nukala[3], Alexander X Gray[2*] and Naga Phani B Aetukuri[1*]

Affiliations:

1. Solid State and Structural Chemistry Unit, Indian Institute of Science, Bengaluru, Karnataka, 560012, India
2. Department of Physics, Temple University, 1925 N. 12th St., Philadelphia, PA 19122, USA
3. Centre for Nano Science and Engineering, Indian Institute of Science, Bangalore, Karnataka, 560012, India
4. Deutsches Elektronen-Synchrotron, DESY, 22607 Hamburg, Germany
5. Utrecht University, Inorganic Chemistry and Catalysis Group Universiteitsweg 99, 3584 CA Utrecht, The Netherlands

*Corresponding authors emails: phani@iisc.ac.in (N.B.A); axgray@temple.edu (A.X.G)

ABSTRACT

*Correlated electron materials (CEMs) host a rich variety of condensed matter phases. Vanadium dioxide ($VO_2$) is a prototypical CEM with a temperature-dependent metal-to-insulator (MIT) transition with a concomitant crystal symmetry change. External control of MIT in $VO_2$ – especially without inducing structural changes - has been a long-standing challenge. In this work, we design and synthesize modulation-doped $VO_2$-based thin film heterostructures that closely emulate a textbook example of filling control in a correlated electron insulator. Using a combination of charge transport, hard x-ray photoelectron spectroscopy, and structural characterization, we show that the insulating state can be doped to achieve carrier densities greater than $5 \times 10^{21}$ $cm^{-3}$ without inducing any measurable structural changes. We find that the MIT temperature ($T_{MIT}$) continuously decreases with increasing carrier concentration. Remarkably, the insulating state is robust even at doping concentrations as high as ~0.2 $e^-$/vanadium. Finally, our work reveals modulation-doping as a viable method for electronic control of phase transitions in correlated electron oxides with the potential for use in future devices based on electric-field controlled phase transitions.*

## INTRODUCTION

Strong electron-electron correlations within narrow d- or f-orbitals underpin a variety of condensed matter phenomena, such as metal-to-insulator transitions (MITs), high-temperature superconductivity, magnetism, and multiferroicity, often observed in correlated electron materials (CEMs).[1,2] $VO_2$ is a prototypical example of a CEM with a temperature-dependent metal-to-insulator phase transition. The electronic phase transition in bulk $VO_2$, which occurs at a MIT temperature ($T_{MIT}$) of ~340 K, is accompanied by a structural phase transition from a metallic rutile phase to an insulating monoclinic phase.[3,4]

The origin of the MIT in $VO_2$ – whether gap-opening is driven by the symmetry-lowering structural transition or by electron-electron correlations - has been widely studied.[5–8] In particular, there is widespread interest in the nature of the insulating state and its external control via doping[9,10], strain[11], oxygen vacancy creation[12], hydrogenation[13], light-and-pulse-induced modulation[14,15], and via electric-fields in a field-effect transistor geometry[16,17].

For example, n-type doping of $VO_2$ with dopants such as $W^{6+}$, $Mo^{5+}$, and $Nb^{5+}$ was shown to decrease $T_{MIT}$ and stabilize the metallic phase.[9,10,18] By contrast, p-type doping of $VO_2$ with dopants such as $Cr^{3+}$, $Ga^{3+}$, and $Al^{3+}$ was shown to increase $T_{MIT}$, thereby stabilizing the insulating phase.[19–21] Similarly, both oxygen vacancy creation and hydrogenation were shown to n-dope $VO_2$ and decrease $T_{MIT}$.[12,13,22] Finally, in $VO_2$ thin films, macroscopic tensile strain along the rutile a-axis was also shown to decrease $T_{MIT}$.[11,12]

In all these previous approaches, modulation of $T_{MIT}$ was always associated with macroscopic changes to the lattice parameters (due to strain) and/or dopant-induced local structural distortions.[9,10,12,13,19,22] In such experiments, where both the lattice strain and carrier concentration change, it is challenging and, sometimes, impossible to disentangle the role of carrier concentration changes from the role of lattice strain. For instance, in the case of W-doped $VO_2$, an increase in W-doping concentration increases both the carrier density and lattice strain.[23]

Other techniques utilizing external stimuli, such as electric-field induced metallization of $VO_2$ in a field-effect transistor geometry, could, in theory, enable the modulation of its conductivity without inducing macroscopic structural changes. However, previous attempts at electric-field-driven metallization of $VO_2$ have not been successful.[12,16,17,24,25] The absence of any field-effect, even when gated through high-K dielectrics, was attributed to the presence of strong correlations in the insulating $VO_2$ phase.[16] Further, ionic-liquid gating of $VO_2$, which could enable accessibility to large interfacial electric-fields, led to oxygen vacancy creation and/or hydrogenation of $VO_2$.[12,16]

Modulation- or remote-doping of oxide semiconductors is an alternative method for achieving high dopant carrier densities without inducing local structural distortions.[26–29] In modulation-doping, a chemical potential mismatch between a high band gap heavily-doped layer (dopant-layer) and a lower band gap undoped layer (channel) leads to charge transfer from the heavily doped dopant-layer to the undoped channel. In general, the dopant layer and the channel are spatially separated by a barrier (or a spacer) layer that kinetically limits the interdiffusion of the dopants from the dopant layer to the channel layer while allowing charge transfer via quantum mechanical tunneling.[30,31]

Modulation-doping was successfully applied to semiconductors and band-insulating oxides such as ZnO and $SrTiO_3$.[27,29,32–35] However, modulation-doping of correlated electron insulators has had limited success. For example, Stemmer and colleagues reported modulation-doping of $NdNiO_3$, but this did not lead to any significant changes in $T_{MIT}$ of the nickelate.[27] Whether modulation-doping could be a generic approach to induce phase transitions in oxides is unclear and several key questions remain unanswered. For example, how do bands evolve in correlated oxides as a function of doping? Can a rigid band model be applied to understand doping in oxides? How sensitive are the ground state properties in correlated oxides to carrier doping?

In this work, we address some of these open questions using the MIT in $VO_2$ as a model system. We propose a modulation-doped heterostructure to n-dope $VO_2$ without inducing any structural distortions. Commonly in modulation-doping, an epitaxial structure is grown with a spacer and the dopant layers epitaxially matched to the semiconducting channel layer. Note that both the spacer and dopant layers must be insulating with a bandgap that is higher than that of the channel layer. However, the only stable rutile oxide that is both insulating with a compatible band mismatch that allows modulation doping of $VO_2$ is $TiO_2$. The other rutile oxides such as $CrO_2$, $RuO_2$, and $IrO_2$ are metallic and therefore not compatible as dopant layers.[36,37]

As an additional challenge, oxygen lattice continuity in epitaxial structures might also lead to oxygen vacancy diffusion across the layers.[22,38,39] We note that oxygen vacancy formation, which was shown to affect the MIT in $VO_2$, is commonly observed in transition metal oxides.[12,40,41] Thus, in order to prevent oxygen vacancy diffusion across the wide band-gap spacer layer as well as to circumvent the lack of lattice-matching insulating rutile oxides, we have gone away from epitaxially-matched modulation-doped heterostructures. Instead, we propose an amorphous $LaAlO_3$ (LAO) layer (with a reported electronic band gap of ~5.6 eV)[42] as the spacer layer. Since LAO has a low oxygen vacancy-diffusivity[43], we use an amorphous oxygen-deficient $TiO_{2-x}$ as the dopant layer. Stoichiometric $TiO_2$ has a bandgap of ~3 eV[44] and $TiO_{2-x}$ is n-type conducting. Using $TiO_{2-x}$ instead of a conventionally doped $TiO_2$ such as, Nb-doped $TiO_2$, significantly simplified heterostructure deposition. Furthermore, this approach avoids the interdiffusion of metallic dopants such as Nb in Nb-doped $TiO_2$ and the associated unintentional doping of $VO_2$.

The modulation-doped structure for all samples used in this work is comprised of a $VO_2$ channel layer, a 2 nm thick LAO spacer layer, and a 3 nm thick $TiO_{2-x}$ dopant layer, as shown schematically in Fig. 1a. All heterostructures were capped with a 1 nm thick LAO layer to prevent dopant passivation from atmospheric impurities as well as oxidation of the $TiO_{2-x}$ dopant layer. Fermi level alignment across the structure is expected to lead to an electron accumulation region at the $LAO/VO_2$ interface. Expected band-alignments for this type-I heterojunction before and after heterostructure formation are shown in Fig. 1b.

To experimentally realize the proposed modulation-doped structure, all samples were grown using pulsed laser deposition (PLD) on single-crystalline $TiO_2$ (001) substrates. $VO_2$ was deposited at 425 °C, while all the other amorphous layers were deposited at room temperature (see methods section for details). It is important to note that room-temperature deposition of the spacer, dopant and capping layers also minimizes any interfacial interdiffusion. A cross-sectional scanning transmission electron microscopy (STEM) image (Fig. 1c) and the associated energy dispersive spectroscopy (EDS) elemental maps (Fig. 1d) show abrupt high-quality interfaces between the $TiO_2$ substrate and the $VO_2$ film, and between the film and LAO spacer. This data is consistent with in-situ RHEED patterns of the deposited $VO_2$ films and suggests that the films are both atomically smooth and single-crystalline (SI Fig. S1). The atomic force microscopy (AFM) images of the complete heterostructures further confirm the high quality of the growth by showing atomically smooth film surfaces (SI Fig. S2).

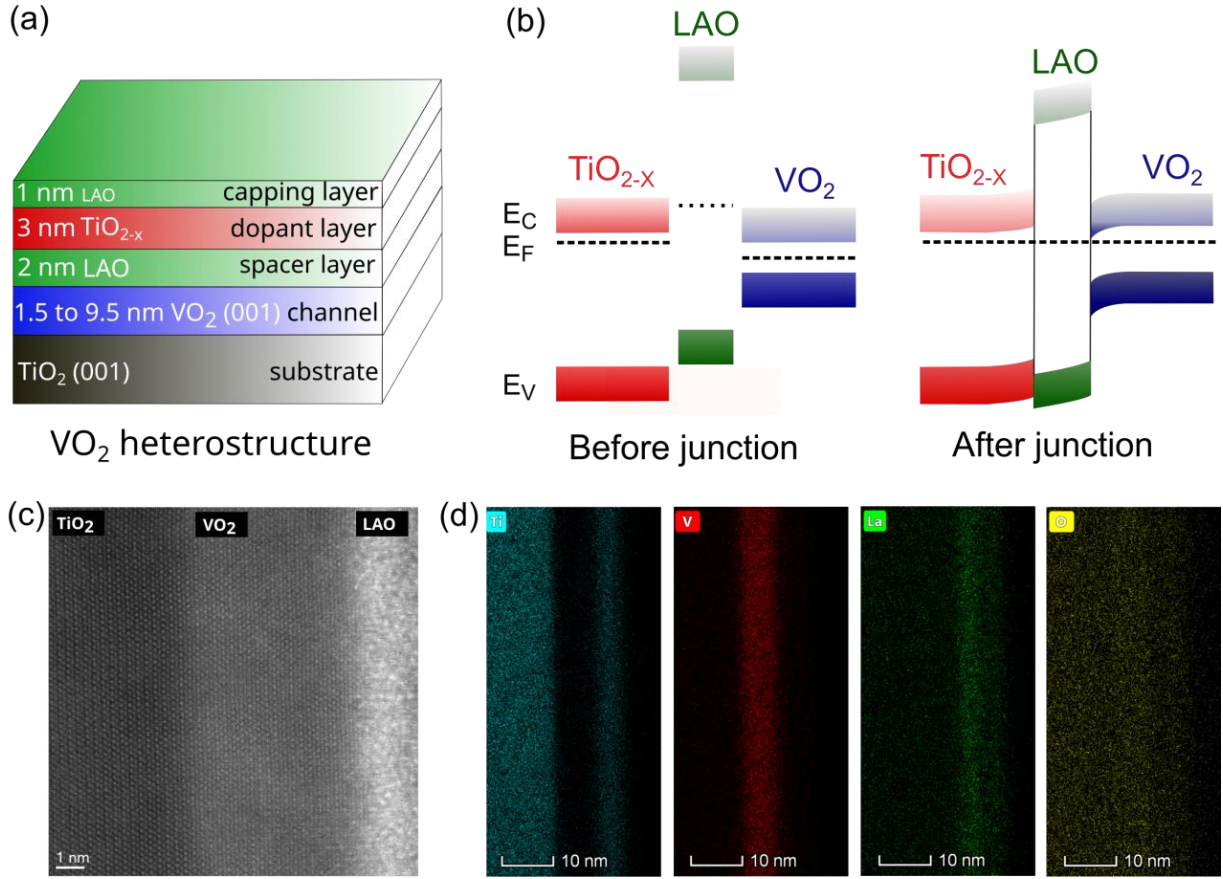

*Fig. 1.* (*a*) *Schematic diagram of the heterostructures used in this work. The thickness of VO$_2$ is varied while the thicknesses of all the other layers are as mentioned in the schematic. (**b**) Schematic energy band diagram for a VO$_2$/LAO/TiO$_{2-x}$ heterostructure before and after the junction formation. Electron accumulation is expected based on the known band offsets. The color intensities are chosen to be proportional to expected electron densities for better visualization. $E_C$, $E_V$, and $E_F$ indicate the conduction band edge, valence band edge, and Fermi level, respectively. (**c**) High-resolution cross-sectional high-angle annular dark-field scanning transmission electron microscopy (HAADF-STEM) image showing abrupt interfaces between TiO$_2$ substrate and VO$_2$ film and VO$_2$ film and the amorphous LAO spacer layer. (**d**) Elemental mapping using energy dispersive x-ray spectroscopy (EDS) showing the various layers in the heterostructure. Note that the scales of (**c**) and (**d**) are different.*

To study the correlation between modulation-doping-induced carrier density changes and the changes in the MIT characteristics, we deposited several modulation-doped heterostructures with varying thicknesses of the VO$_2$ layer ranging from 1.5 nm to 9.5 nm, while keeping the thickness of the TiO$_{2-x}$ layer unchanged at 3 nm. Thomas-Fermi screening lengths in VO$_2$ are expected to be on the order of 1 nm (SI section S3) and, therefore, the highest n-type carrier densities are expected for the lowest VO$_2$ film thickness used in this study (~1.5 nm).

Heterostructures with VO$_2$ films thinner than ~1.5 nm were not attempted due to the expected titanium interdiffusion at the VO$_2$ film and TiO$_2$ substrate interface.[45,46] We note that interfacial titanium interdiffusion will be present in thicker films as well. However, at VO$_2$ thicknesses greater than 1.5 nm, there is still an observable MIT.

A summary of the θ-2θ X-ray diffraction (XRD) measurements, performed at room temperature, for a 9.5 nm VO$_2$ film and the VO$_2$/LAO/TiO$_{2-x}$/LAO heterostructures on TiO$_2$(001) substrates are shown in Fig. 2a. Clearly, the angular positions of the VO$_2$ (002) Bragg reflection peaks are identical for both the 9.5 nm VO$_2$ film (purple spectrum) and the 9.5 nm VO$_2$ heterostructure (blue spectrum). Furthermore, it is evident that the angular position of the Bragg reflection is essentially independent of the thickness of the VO$_2$ film in the heterostructure. Additionally, there were no significant changes in the θ-2θ X-ray diffractograms between VO$_2$ films and heterostructures with the same VO$_2$ thickness (SI Fig. S4). Reciprocal space maps also confirm that all samples are coherently strained in the plane of the TiO$_2$(001) substrate. The out-of-plane rutile c-axis lattice parameter is identical for the thin film and the heterostructures for all thicknesses of VO$_2$ (SI Fig. S5). Based on the θ-2θ XRD measurements, reciprocal space maps, and cross-sectional STEM imaging we conclude that the lattice parameter changes, if any, are within the instrumental resolution (better than 0.1 pm) for all the VO$_2$ heterostructures used in this work. We also note that the reflections for the LAO spacer and capping layers and the TiO$_{2-x}$ dopant layer are absent, suggesting that these layers are not crystalline.

Next, we discuss the variations in the MIT characteristics for the same set of samples as used in the XRD studies. As shown in Fig. 2b, $T_{MIT}$ systematically decreases with decreasing VO$_2$ thickness in the heterostructure. Note that the decrease in $T_{MIT}$ for thin films of VO$_2$ is thickness independent (SI Fig. S6), suggesting that the decrease in $T_{MIT}$ for VO$_2$ heterostructures is intrinsic to heterostructure formation. Furthermore, the sheet resistance of VO$_2$ heterostructures in the insulating state also decreases (SI Fig. S7). Except for the heterostructure with the thinnest VO$_2$ layer (1.5nm), all films continue to show a positive temperature coefficient of resistance, suggesting metallicity above $T_{MIT}$.

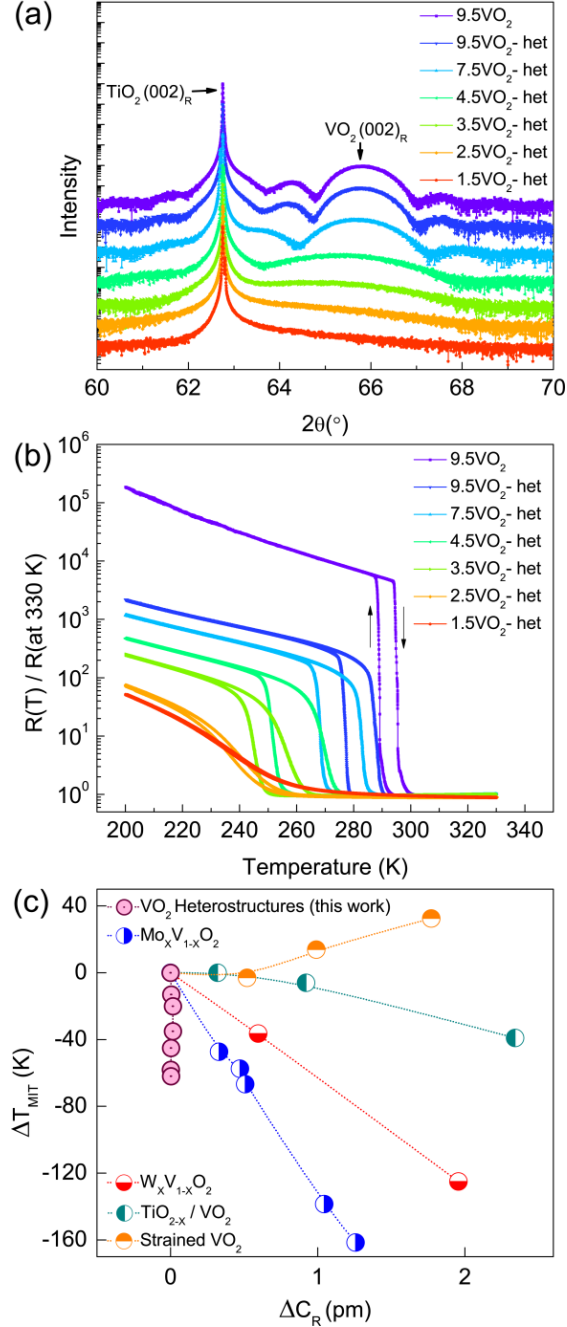

*Fig. 2. (a) High-resolution θ-2θ XRD spectra for the 9.5 nm VO$_2$ thin film and for VO$_2$ heterostructures with varying thicknesses, t. For nomenclature simplicity, we distinguish VO$_2$ thin films and heterostructures with a VO$_2$ thickness of 't' as tVO$_2$ and tVO$_2$-het, respectively. For example, 9.5VO$_2$-het corresponds to a heterostructure with 9.5 nm thick VO$_2$. (b) Resistance versus temperature plots for the same set of samples as shown in (a). Resistance values presented here are normalized to the resistance at 330 K and were measured in Van der Pauw geometry (also see SI Fig. S3). Clearly, $T_{MIT}$ decreases with decreasing VO$_2$ thickness in the heterostructure. (c) A comparison of the changes in $T_{MIT}$ versus the changes in the rutile C-axis lattice parameter ($\Delta T_{MIT}$ vs $\Delta C_R$) for this work and other previously published work using W[9]- and Mo[10]-doping (red and blue respectively), oxygen vacancy doping[22] (gray) and strain[11] (orange). The relative changes are compared to the undoped and unstrained states in the case of bulk doping and for strained VO$_2$. For the modulation-doped heterostructures, the reference state is taken as a 9.5 nm VO$_2$ thin film on TiO$_2$ (001) substrate.*

We summarize our observations and compare the changes in $T_{MIT}$ with other previously published studies in Fig. 2c (also see SI Fig. S8 for details regarding $T_{MIT}$ calculation). Significantly, there is a nearly 60 K change in $T_{MIT}$ for the thinnest heterostructures without any measurable changes to the rutile C-axis. In contrast, any comparable change in $T_{MIT}$ in the literature is associated with $\Delta C_R$ greater than 0.5 pm. This demonstrates control over the MIT in $VO_2$ without any measurable structural changes in the $VO_2$ heterostructures proposed and synthesized in this work. We note that a decrease in $T_{MIT}$ was observed for elemental doping of $VO_2$ with n-type dopants such as W and Mo, while an increase in $T_{MIT}$ was observed for hole-doping with elements such as Cr and Al. There is no W or Mo in any of the heterostructures in this work, and both La and/or Al doping can be ruled out because they would (if anything) lead to hole-doping resulting in an increase in the $T_{MIT}$. This is contrary to the decrease in $T_{MIT}$ observed in these $VO_2$ heterostructures.

To measure the extent and type of doping, we performed temperature-dependent Hall measurements. Hall measurements show an enhancement in the carrier densities in the insulating state with the Hall coefficient indicative of electron-doping (Fig. 3a). On the other hand, carrier densities in the insulating phase increased from $\sim 6 \times 10^{17}$ cm$^{-3}$ (for 9.5 nm $VO_2$ thin film) to $\sim 2.8 \times 10^{19}$ cm$^{-3}$ (for 9.5 nm $VO_2$ heterostructure) to a highest of $\sim 5 \times 10^{21}$ cm$^{-3}$ for the 2.5 nm $VO_2$ heterostructure. Contrastingly, the metallic state carrier densities remain identical ($\sim 6 \times 10^{22}$ cm$^{-3}$) across all the $VO_2$ heterostructures and are consistent with previous reports.[12,24]

However, carrier mobility decreases in both the insulating and metallic states as the thickness of the $VO_2$ layer decreases (Fig. 3b). In the metallic state, this is potentially due to contributions from interfacial scattering, which increases with decreasing film thickness. In the insulating state, the decrease in carrier mobility could result in part from increased electron-electron scattering and interfacial scattering. A summary of the changes in carrier concentration is plotted against $T_{MIT}$ in Fig. 3c. There is a clear correlation between the $T_{MIT}$ and the carrier density with the highest carrier density of $\sim 5 \times 10^{21}$ cm$^{-3}$ stabilizing the metallic state of $VO_2$ to a $T_{MIT}$ of $\sim 237$ K (SI Fig. S8). Importantly, the continuous increase in carrier density with decreasing $VO_2$ thickness without any lattice parameter changes is suggestive of modulation-doping in $VO_2$ heterostructures.

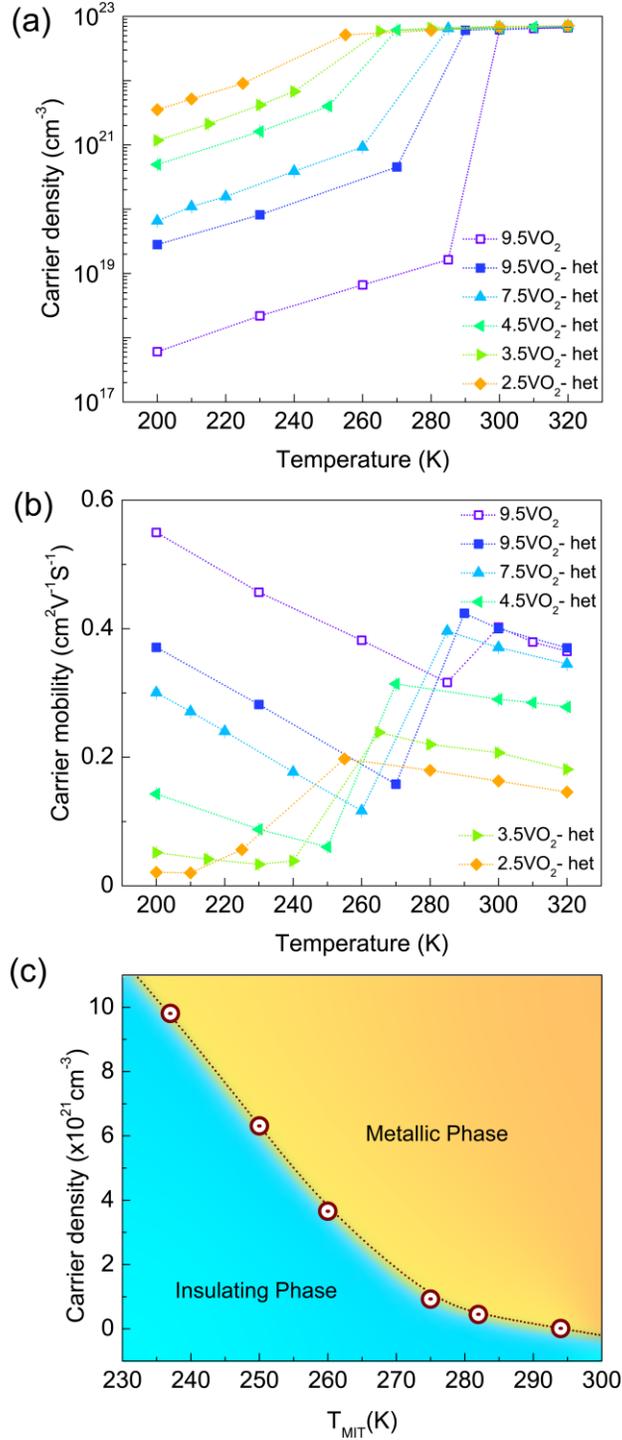

***Fig. 3.*** *Plots of temperature-dependent (**a**) carrier densities and (**b**) carrier mobilities for tVO$_2$ and tVO$_2$-het samples as mentioned in the legends. Carrier density in the insulating state increases with decreasing VO$_2$ thickness while carrier mobility decreases. (**c**) A phase diagram from the results in (**a**). The dotted lines connected across the data points are guide to the eye.*

To further establish that most of the carriers are induced by modulation doping, we prepared two additional control samples. The first is a 7.5 nm $VO_2$ thin film with a 2 nm LAO cap layer (7.5$VO_2$-LAO) and the second is a 7.5 nm $VO_2$ heterostructure with a 3 nm stoichiometric $TiO_2$ layer (7.5$VO_2$-LAO-$TiO_2$). We compared the MIT characteristics of these two samples with the MIT characteristics of the 7.5 nm $VO_2$ thin film (7.5$VO_2$) and a 7.5 nm $VO_2$ heterostructure with a $TiO_{2-x}$ dopant layer (7.5$VO_2$-het). A summary of sheet resistance versus temperature data is shown in SI Fig. S9. The 7.5 nm $VO_2$ heterostructure with a $TiO_{2-x}$ dopant layer has the lowest sheet resistance and the lowest $T_{MIT}$ with a $T_{MIT}$ change of ~20 K. By contrast, the decrease in $T_{MIT}$ was restricted to ~6 K after the deposition of the 2 nm LAO layer. Remarkably, there is no further decrease in $T_{MIT}$ in the heterostructure with stoichiometric $TiO_2$ layer, suggesting that the $TiO_{2-x}$ dopant layer is required for the observed large change in $T_{MIT}$.

Consistent with this, the carrier density in the 7.5 nm $VO_2$ heterostructure with the $TiO_{2-x}$ dopant layer is ~$7 \times 10^{19}$ cm$^{-3}$ compared to a carrier density of $2 \times 10^{19}$ cm$^{-3}$ for the 7.5 nm $VO_2$ capped with 2 nm LAO (SI Fig. S10). It is possible that amorphous (disordered) LAO could host ionized donors and enable modulation-doping.[29] However, we found that the amorphous LAO deposited for these experiments is insulating, suggesting that any ionized donors should be below the measurement threshold of electrical resistivity measurements. We estimate that such ionized donors in LAO, if any, should have a carrier density of ~$10^{19}$ cm$^{-3}$ (assuming a carrier mobility of 0.01 cm$^2$/V-s) or lower, putting an upper bound on the number of carriers contributed by the spacer layer.

To probe the electronic band bending that enables electron accumulation in the $VO_2$ channel layer, we performed bulk-sensitive hard X-ray photoelectron spectroscopy (HAXPES)[47] measurements at the P22 beamline in the PETRA III synchrotron at DESY. We note that standard ultra-violet photoemission (UPS) or soft X-ray photoemission measurements cannot facilitate a probing depth sufficient to reach the $VO_2$/LAO interface that is buried beneath multiple layers of the heterostructure. To capture interfacial band bending in $VO_2$, HAXPES measurements were performed in both the insulating phase (at 200 K) and the metallic phase (at 310 K) for $VO_2$ heterostructures with $VO_2$ thicknesses of 1.5 nm, 3.5 nm, 4.5 nm, and 7.5 nm, and for a $VO_2$ thin film with a thickness of 7.5 nm as a reference.

For the 7.5 nm $VO_2$ film measured at 200 K (insulating state), the binding energies of the V $2p_{3/2}$ and V $2p_{1/2}$ core-level peaks were observed to be ~515.8 eV and ~523.1 eV, respectively (see Fig. 4a). These measured binding energies (see SI Fig. S11 for binding energy calibration procedure) are consistent with previous reports.[48–50] Importantly, a systematic shift of the main component of the V $2p_{3/2}$ peak to higher binding energies is observed for the $VO_2$ heterostructures, with the highest increase in binding energy (~250 meV) observed for the heterostructure with the thinnest $VO_2$ layer (1.5 nm), as seen in the inset. This is also in agreement with the highest carrier densities and the lowest $T_{MIT}$ being observed for the heterostructures with the thinnest $VO_2$ layers. For measurements performed on $VO_2$ in the metallic state, no such binding energy shift was observed (Fig. 4b and SI Fig. S12). This is consistent with the complete screening of interfacial electric fields at the metallic $VO_2$/LAO interface (Figs. 4c and d). The presence of binding energy shifts observed in the insulating state of $VO_2$ and their absence in the metallic state of $VO_2$ further support carrier doping by chemical potential shifts (modulation-doping) in the insulating state for $VO_2$ heterostructures as proposed in this work.

Photoemission data also showed two remarkable features in the V $2p$ spectra. The first, labelled *P1* in Figs. 4a and 4b, is a lower binding energy shoulder around ~514.5 eV in the insulating and metallic states. The presence of this spectral feature at lower binding energies was proposed to signify non-local screening from coherent $3d^1$ states near $E_F$ in the metallic phase of $VO_2$.[50,51] Interestingly, the intensity of *P1* in the insulating state increases with increasing carrier density and decreasing $VO_2$ thickness in the heterostructures. The emergence of this peak in the insulating state spectra for $VO_2$ heterostructures suggests that the additional charge transferred to the $VO_2$ channel layer enables non-local screening that was previously observed only in the metallic phase of $VO_2$.

To further quantify the evolution of the *P1* peak across the metallic and insulating phases, we compared the metallic and insulating state spectra of 1.5 nm and 7.5 nm $VO_2$ heterostructures in SI Fig. S13. The intensity of the *P1* peak in the insulating state of the 7.5 nm $VO_2$ heterostructure is observable but small in comparison to the *P1* peak in the metallic state (SI Fig. S13a). The difference spectrum shows a large difference at ~514.5 eV (SI Fig. S13c) further confirming that the non-locally screened shoulder is negligibly small in the insulating state when compared to the metallic state. Remarkably, the non-locally screened shoulder is quite predominant in the insulating state spectra of 1.5 nm VO$_2$ heterostructure (SI Fig. S13b). The intensity difference spectra between the metallic and insulating states shows a very small difference between the two phases (Fig. S13d). These trends are consistent with the differences in the carrier densities between the metallic and insulating phases. The carrier density ratio between the metallic and insulating states is about 10 for the 1.5 nm VO$_2$ heterostructure compared to about 1000 for the 7.5 nm VO$_2$ heterostructure.

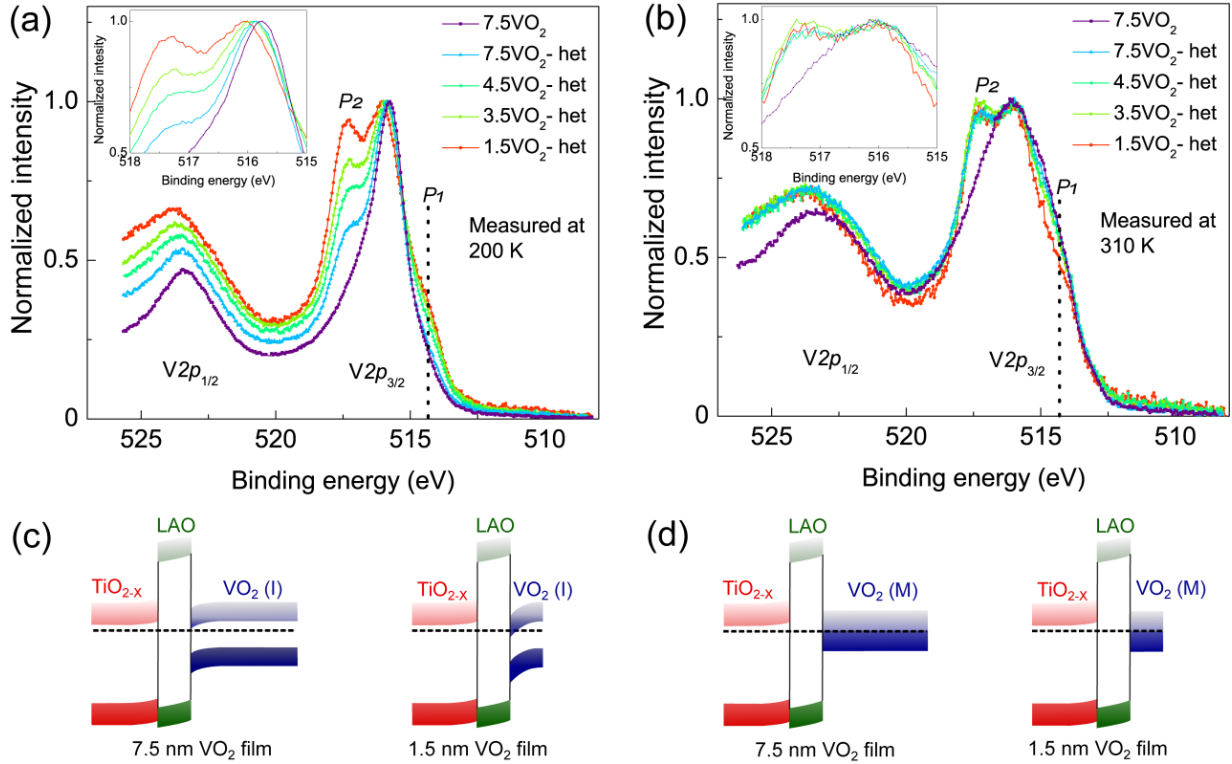

*Fig. 4.* A comparison of V 2p core-level spectra of modulation doped VO$_2$ heterostructures for (**a**) the insulating (200 K) and (**b**) the metallic states (at 310 K). A clear shift in the V 2p levels is seen in the insulating state spectra but not in the metallic state spectra. Schematics in (**c**) and (**d**) show the expected band-bending in the modulation-doped heterostructures for the insulating and metallic states respectively. Band-bending is expected in the insulating state and not in the metallic state. Two additional spectral features not seen in VO$_2$ thin films are labelled P1 and P2.

The second remarkable feature in the photoemission data is a high binding energy spectral feature at ~517.5 eV. This feature is associated with the V $2p_{3/2}$ peak and labelled *P2* in Figs. 4a and b. A corresponding feature is also observed for the V $2p_{1/2}$ peak but is more smeared out and appears as broadening on the higher-binding-energy side at ~525 eV. In general, higher binding energy spectral features are associated with higher oxidation states in photoemission. The presence of $V^{5+}$ is a possibility. However, an increase in the oxidation state from $V^{4+}$ to $V^{5+}$ cannot explain the observed electron-doping in $VO_2$ heterostructures. Generally, electron doping should decrease the $V^{4+}$ oxidation state in $VO_2$ and therefore, an increase in the oxidation state of vanadium cannot explain the increase in electron density in the $VO_2$ heterostructures. Therefore, $V^{5+}$, even if present, has no bearing on the MIT observed in heterostructures.

*P2* was also observed in $VO_2$ samples capped with 2 nm LAO (SI Fig. S14). Therefore, we have also inspected the La $3d_{5/2}$ and Al $1s$ spectra to look for any chemical shifts associated with a redox or chemical reaction at the $VO_2$/LAO interface. As shown in SI Fig. S15, there are no observable changes to the spectra across heterostructures. Finally, interfacial oxygen vacancy creation remains a possibility. However, any oxygen vacancy creation should lead to $V^{3+}$ and an associated low-binding-energy feature in both the metallic and insulating state spectra. However, the spectra do not show any signatures of oxygen vacancies in $VO_2$ heterostructures. Therefore, we rule out any oxygen-vacancy induced carrier doping in these heterostructures.

Furthermore, the intensity of *P2* is carrier density dependent. For all V $2p$ spectra in the metallic state, where the carrier density is independent of the $VO_2$ thickness, the intensity of this additional peak relative to the main V $2p_{3/2}$ peak is also independent of the $VO_2$ thickness, with the intensity ratio of *P2* to V $2p_{3/2}$ being close to 1. Contrastingly, the intensity of *P2* increases with the decreasing film thickness in the insulating state of $VO_2$. The intensity ratio of *P2* to V $2p_{3/2}$ approaches the intensity ratio observed for the metallic state spectra at the highest carrier density of $\sim 5 \times 10^{21}$ cm$^{-3}$ in the insulating state. These carrier-density-dependent changes suggest that this new spectral feature is intrinsic to the heterostructure. However, this additional spectral feature might benefit from further spectroscopic investigation with complementary techniques such as X-ray absorption spectroscopy to confirm its origins.

The combination of electron transport and HAXPES data show that $VO_2$ heterostructures facilitated effective modulation doping and carrier densities as high as $5 \times 10^{21}$ cm$^{-3}$ could be achieved using this approach. The highest carrier densities correspond to electron doping of $\sim 0.2$ e-/vanadium. This is an extremely high dopant density at which conventional rigid band models predict metallization in doped correlated insulators.[52]

Bulk-sensitive valence-band HAXPES spectra recorded for the same set of heterostructures as discussed in Fig. 4 are shown in Fig. 5 and SI Fig. S16. The insulating-state spectra for all $VO_2$ heterostructures (blue) exhibit nearly zero spectral intensity at the Fermi level while an appreciable non-zero spectral intensity is observed for the higher-temperature metallic-state spectra (orange). These spectra further confirm that $VO_2$ continues to undergo a MIT even in the presence of electron densities as high as ~0.2 e-/vanadium. The presence of MIT at such high doping levels, without any observable changes in the lattice parameters (Fig. 2a and SI Fig. S4), points to a possible renormalization of the electronic structure with doping.

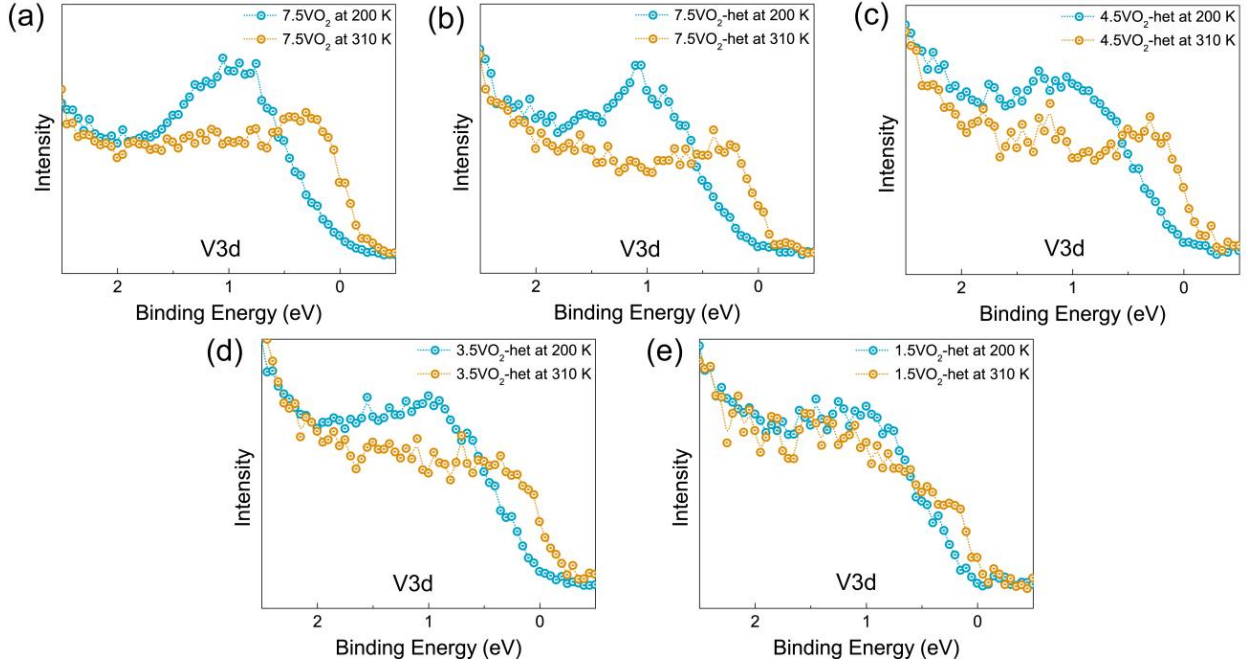

*Fig. 5. A comparison of the V 3d valence band (VB) spectra of modulation-doped $VO_2$ heterostructures for the insulating (200 K, blue) and metallic states (at 310 K, orange) for different $VO_2$ film thicknesses of (**a**) 7.5 nm $VO_2$ film and (**b**) 7.5 nm, (**c**) 4.5 nm, (**d**) 3.5 nm, and (**e**) 1.5 nm $VO_2$ heterostructures. There is a clear spectral weight shift across the MIT for all the samples with the insulating state being robust even for the heterostructure with a $VO_2$ thickness of 1.5 nm, corresponding to carrier doping of ~0.2 e-/Vanadium.*

## Conclusions:

In summary, we demonstrated a purely electronic control of the MIT in modulation-doped $VO_2$ heterostructures. Our work shows that the insulating state in $VO_2$ is surprisingly robust even in the presence of electron doping as high as 0.2 e-/vanadium. Notably, all the films meet the Mott criterion ($a_B \cdot n_C^{\frac{1}{3}} > 0.25$, where $a_B$ is the effective Bohr radius and $n_C$ is the carrier density). Therefore, metallicity is expected at all temperatures based on the carrier densities achieved in these experiments. We note that a similar robust insulating state had also been found in modulation-doped nickelate thin films.[27] Perhaps, the development of theoretical models that go beyond the conventional carrier concentration independent rigid-band models will be required to understand electronic phase transitions in correlated electron oxides. An alternate explanation could be that the insulating state is favored at lower thicknesses due to interfacial disorder-induced Anderson-like localization of carriers, which will be more pronounced for the thinnest films. Further experiments will be needed to assess whether the insulating state is stabilized by the interfacial disorder.

A remarkable feature of this work is the possibility of bulk metallization in modulation-doped $VO_2$. In general, in band semiconductors such as $SrTiO_3$, conductivity modulation is achieved over a thickness of 1-2 nm in the vicinity of the channel/spacer interface.[29,53] In this work, a sharp MIT is observed for heterostructures with $VO_2$ thicknesses as high as 9.5 nm, which are much higher than the Thomas Fermi screening length of ~1 nm. This is suggestive of the entirety of the film being metallized at the lowered $T_{MIT}$ after modulation doping, even though the charge transfer densities are the highest at the interface (within 1-2 nm of the interface). While further studies are required to establish this beyond doubt, interfacial-doping induced bulk metallization of correlated electron insulators has implications for low-power electronics with high ON/OFF ratios.[28,54]

Finally, our work shows that modulation doping is a powerful technique for achieving high carrier densities, close to those possible with elemental doping. Since our approach does not need any epitaxially matched spacer and dopant layers, it expands the library of materials that can be explored for the study of modulation-doping-induced electronic phase transitions of other related CEMs including complex oxides[2] and pyrochlores[55]. This methodology, therefore, paves the way for exploring 'pure' electronic effects in correlated oxides and related systems. Such studies could also enable a fundamental understanding of band matching and relevant energy scales in complex oxides and, perhaps enable the discovery of new interfacial phases and devices that rely on phase transitions.

## Methods

Prior to deposition, single-crystalline rutile $TiO_2$ (001) substrates (Shinkosa, Japan) were treated using the procedure discussed previously.[56] All thin film samples were deposited using PLD (NEOCERA) with a 248 nm KrF laser. All $VO_2$ thin films (both pristine films and heterostructures) were deposited on treated $TiO_2$ substrates from a sintered $V_2O_5$ target with a laser fluence of ~1.5 J/cm$^2$, 8 mTorr of oxygen pressure, and a growth rate of ~4.7×10$^{-2}$ Å/pulse at a substrate temperature of 425 °C.[56] For all heterostructure samples, 2 nm thick LAO spacer layers were deposited at 10 mTorr of $O_2$ pressure at a growth rate of ~5×10$^{-2}$ Å/pulse using a single-crystalline LAO target (Shinkosa Japan). 3 nm thick $TiO_{2-x}$ dopant layers were then deposited using a $TiO_{2-x}$ single-crystalline target (Shinkosa Japan) at a growth rate of ~4.2×10$^{-2}$ Å/pulse in 10$^{-5}$ Torr of background vacuum. Finally, a 1 nm thick LAO layer was deposited under the same conditions used for the LAO spacer layer. Depositions of the spacer, dopant, and capping layers were all done at room temperature at a laser fluence of ~1.2 J/cm$^2$. For all depositions, the substrate-to-target distance was maintained at 55 mm.

High-resolution Cu-$K\alpha$ X-ray diffraction spectra for both the pristine and heterostructure $VO_2$ films were recorded in standard θ–2θ geometry using a Rigaku Smart Lab X-ray diffractometer. LEPTOS 7.8 software by Bruker was used to determine film thicknesses and used to calculate the differential strain between pristine and heterostructure films.

Cross-sectional scanning transmission electron microscopy (STEM) imaging and electron dispersive x-ray spectroscopy (EDS) mapping were performed using TITAN Themis microscope (60-300 kV) equipped with a probe corrector and super-X four quadrant EDS detector. The high angle annual dark field (HAADF)-STEM images were acquired at an operating potential of 300 kV with a convergence angle of 24.5 mrad, 160 mm camera length, and a dwell time of 12 μs per pixel. The images were further processed with Gatan digital micrograph software. The EDS maps were acquired using Velox software under similar microscopic conditions with a dwell time of 2 μs per pixel.

Sheet resistance vs temperature measurements were performed in Van der Pauw geometry using Keithley 2450 SMU and Eurotherm 2408 PID temperature controller. Continuous temperature scanning was carried out at a rate of 4 K/minute for both heating and cooling cycles.

To extract carrier density and mobility, Hall measurements for all films and heterostructures were performed using Van der Pauw geometry in a PPMS-Dynacool equipment from Quantum Design and Keithley SMU 2450 from Tektronix. For these measurements, the magnetic field was swept from 0 T to 2 T to -2 T to 0 T at a scan rate of 100 Oe/s for different temperatures ranging from 200 K to 320 K.

High-resolution hard X-ray photoelectron spectroscopy (HAXPES) measurements[57] were carried out with an incident photon energy of 6.2 keV at the sample temperatures of 200 K (insulating phase) and 310 K (metallic phase). Binding energy calibration was carried out using a high-resolution Fermi-edge measurement on a standard Au sample. Core-level and valence-band spectra were measured using a wide acceptance angle SPECS Phoibos 225HV hemispherical electrostatic analyzer in a near-normal emission geometry. The total energy resolution was estimated to be approximately 320 meV. Preliminary HAXPES measurements and sample screening were carried out using a lab-based HAXPES instrument at Temple University equipped with a wide acceptance angle ScientaOmicron EW4000 analyzer at a total experimental energy resolution of 450 meV.

**Acknowledgments:**
AXG, AMD, and JRP acknowledge support from the DOE, Office of Science, Office of Basic Energy Sciences, Materials Sciences, and Engineering Division under Award No. DE-SC0019297. The electrostatic photoelectron analyzer for the lab-based HAXPES measurements at Temple University was acquired through an Army Research Office DURIP grant (Grant No. W911NF-18-1-0251). We acknowledge DESY (Hamburg, Germany), a member of the Helmholtz Association HGF, for the provision of experimental facilities. Beamtime at DESY was allocated for proposal I-20210142. Funding for the HAXPES instrument at beamline P22 by the Federal Ministry of Education and Research (BMBF) under framework program ErUM is gratefully acknowledged. PN and RKR acknowledge the Advanced Facility for Microscopy and Microanalysis (AFMM) for providing the electron microscope and FIB facility. A.N. acknowledges support from the startup grant at Indian Institute of Science (SG/MHRD-19-0001). The authors acknowledge CeNSE, IISc for access to HR-XRD, wire bonding, and clean-room facilities. N.B.A. acknowledges the new faculty startup grant provided by the Indian Institute of Science under Grant No. 12-0205-0618-77. N.B.A. is thankful to Professor Anil Kumar for access to the PLD system. S.R.M. and D.M. want to thank Jibin J. Samuel and Mithun Ghosh for useful discussions. We thank Professor Satish Patil for providing access to facilities supported by the Swarnajayanti fellowship under Grant No. DST/SJF/CSA-01/2013-14. AFM measurements were performed on a Cypher-ES AFM funded by the DST-FIST program and Hall measurements were performed on a PPMS-Dynacool system funded under the UGC-CAS program.

# Supplementary Information

# Modulation-Doping a Correlated Electron Insulator

Debasish Mondal[1], Smruti Rekha Mahapatra[1], Abigail M Derrico[2], Rajeev Kumar Rai[3], Jay R Paudel[2], Christoph Schlueter[4], Andrei Gloskovskii[4], Rajdeep Banerjee[1], Frank M F DeGroot[5], Dipankar D Sarma[1], Awadhesh Narayan[1], Pavan Nukala[3], Alexander X Gray[2*] and Naga Phani B Aetukuri[1*]

Affiliations:

1. Solid State and Structural Chemistry Unit, Indian Institute of Science, Bengaluru, Karnataka, 560012, India
2. Department of Physics, Temple University, 1925 N. 12th St., Philadelphia, PA 19122, USA
3. Centre for Nano Science and Engineering, Indian Institute of Science, Bangalore, Karnataka, 560012, India
4. Deutsches Elektronen-Synchrotron, DESY, 22607 Hamburg, Germany
5. Utrecht University, Inorganic Chemistry and Catalysis Group Universiteitsweg 99, 3584 CA Utrecht, The Netherlands

*Corresponding authors emails: phani@iisc.ac.in (N.B.A); axgray@temple.edu (A.X.G)

**Table of Contents:**

## S1: Reflection High Energy Electron Diffraction (RHEED) of VO$_2$ thin films

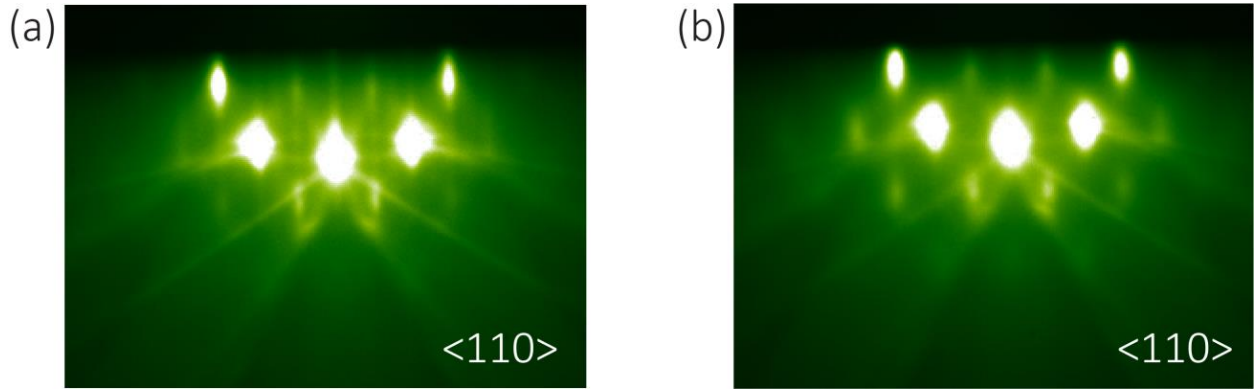

***Fig. S1.*** *RHEED pattern of VO$_2$ thin films on TiO$_2$ (001) substrate along the <110> direction for two different thicknesses of (**a**) 1.5 nm and (**b**) 9.5 nm. These patterns were captured at 425 °C just after the VO$_2$ deposition. RHEED patterns are indicative of single-crystalline VO$_2$ films with smooth film surfaces.*

## S2: AFM images of VO$_2$ thin film and heterostructure

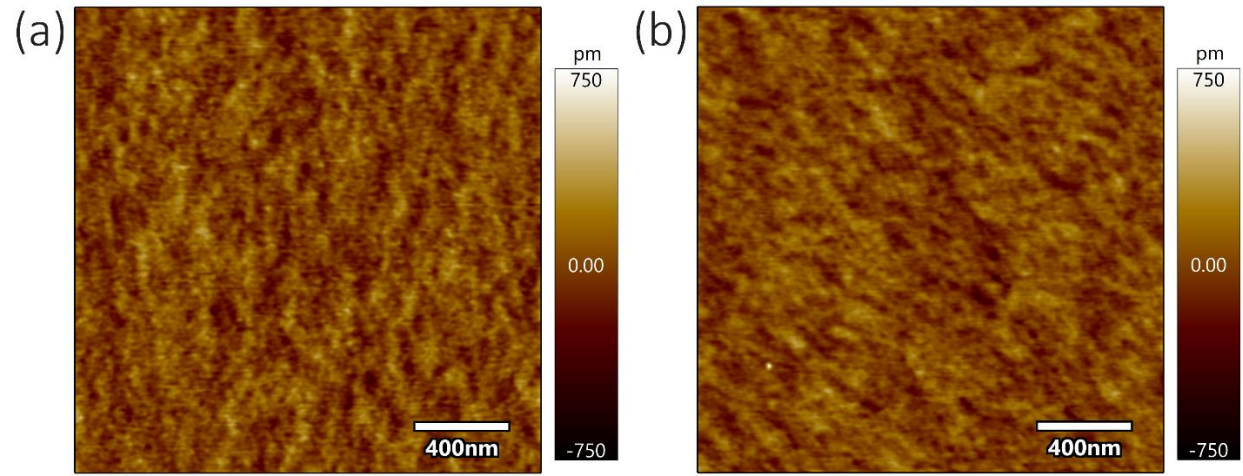

***Fig. S2.*** *AFM images of (**a**) 9.5 nm VO$_2$ film and (**b**) 9.5 nm VO$_2$ heterostructure on TiO$_2$ (001) substrate show smooth 2D surfaces with a root mean-square roughness in the range of 80-100 pm. The images were taken in the tapping mode (AC air topography) using an Asylum Cypher ES AFM.*

## S3: Thomas-Fermi screening length calculation for VO₂ heterostructures

Thomas-Fermi screening length (L) can be calculated as follows

$$L = \sqrt{\frac{K\varepsilon_0 T k_B}{e^2 n_e}}$$

where critical carrier density, $n_e \approx \left(\frac{0.25}{a_B}\right)^3$ and $K, \varepsilon_0, k_B, T, e, a_B$ are the dielectric constant ($\approx$ 36)[1,2] of VO₂, free space permittivity ($\approx$ 8.854x10⁻¹² F/m ), Boltzmann constant ($\approx$ 1.380x10⁻²³ m² kg s⁻² K⁻¹), temperature of VO₂, electronic charge ($\approx$ 1.602x10⁻¹⁹ C) and effective Bohr radius respectively. Effective Bohr radius ($a_B$) can be calculated as follows

$$a_B = \frac{h^2 K \varepsilon_0}{\pi m^* e^2}$$

where $h$ is the Planck's constant ($\approx$ 6.626x10⁻³⁴ m²kg/s) effective mass of electron in VO₂, $m^* \approx 3.5 m_e$.[2,3] By plugging these above values and keeping T $\approx$ 300 K, calculated Thomas-Fermi screening length (L) is ~0.73 nm.

## S4: A comparison of X-ray diffractograms for VO₂ thin films and heterostructures

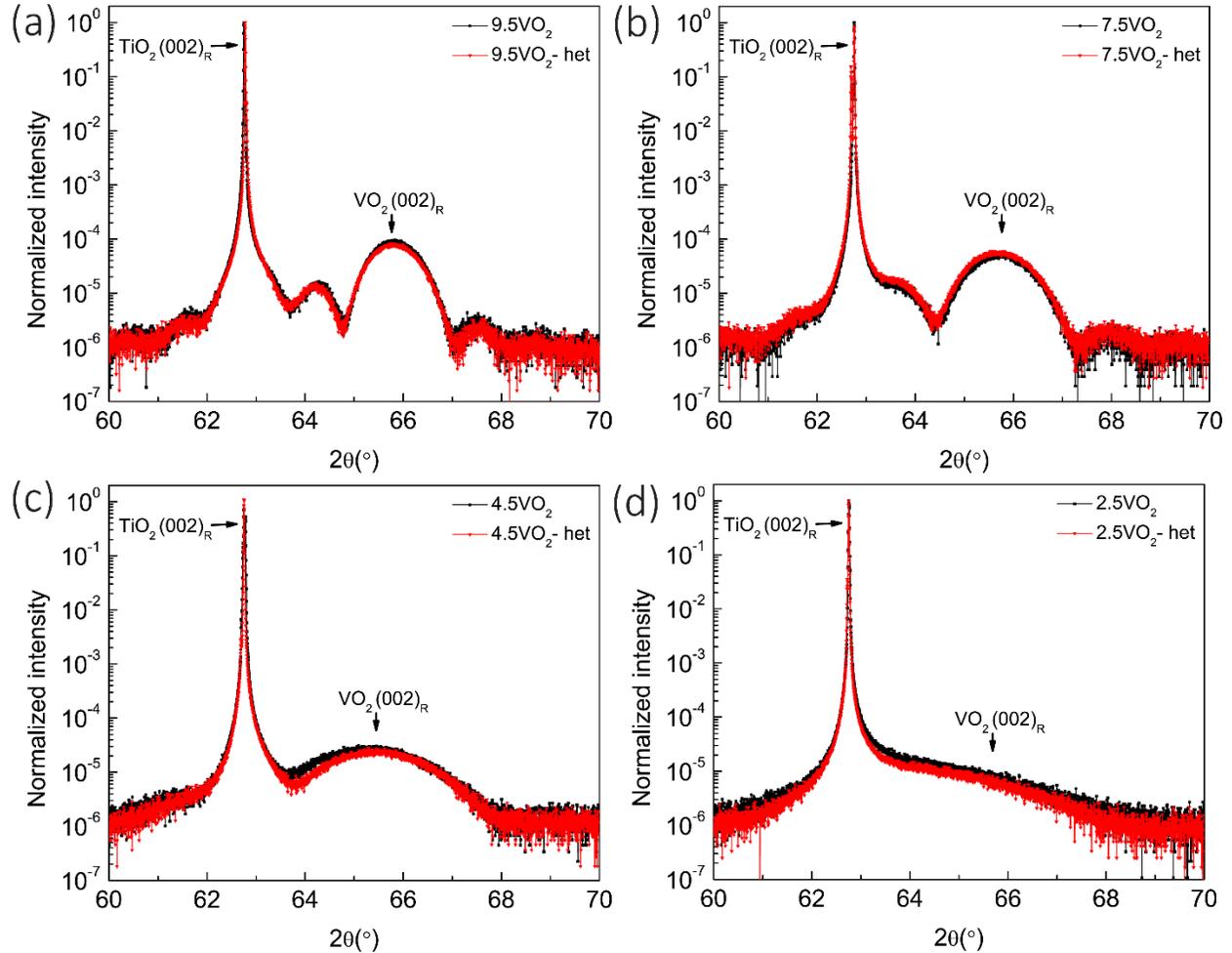

***Fig. S4.*** *Comparison of high-resolution θ-2θ X-ray diffractograms of VO₂ (001) thin films (black) and heterostructures (red) at different VO₂ thicknesses of (**a**) 9.5 nm, (**b**) 7.5 nm, (**c**) 4.5 nm and (**d**) 2.5 nm. We showed in Fig. 2a of the main manuscript that there is no measurable change in the VO₂ lattice parameter, $C_R$, between VO₂ thin films and heterostructures. To further confirm this, we performed and compared high-resolution X-ray diffractograms for thin films and heterostructures with identical VO₂ thickness. Clearly, there is excellent overlap of the two diffractograms, including thickness fringes, for all thicknesses measured. This is clear indication that depositing heterostructures on ultra-thin VO₂ films did not affect the lattice parameters of VO₂.*

.

## S5: Reciprocal space maps (RSM) of VO$_2$ thin film and heterostructures

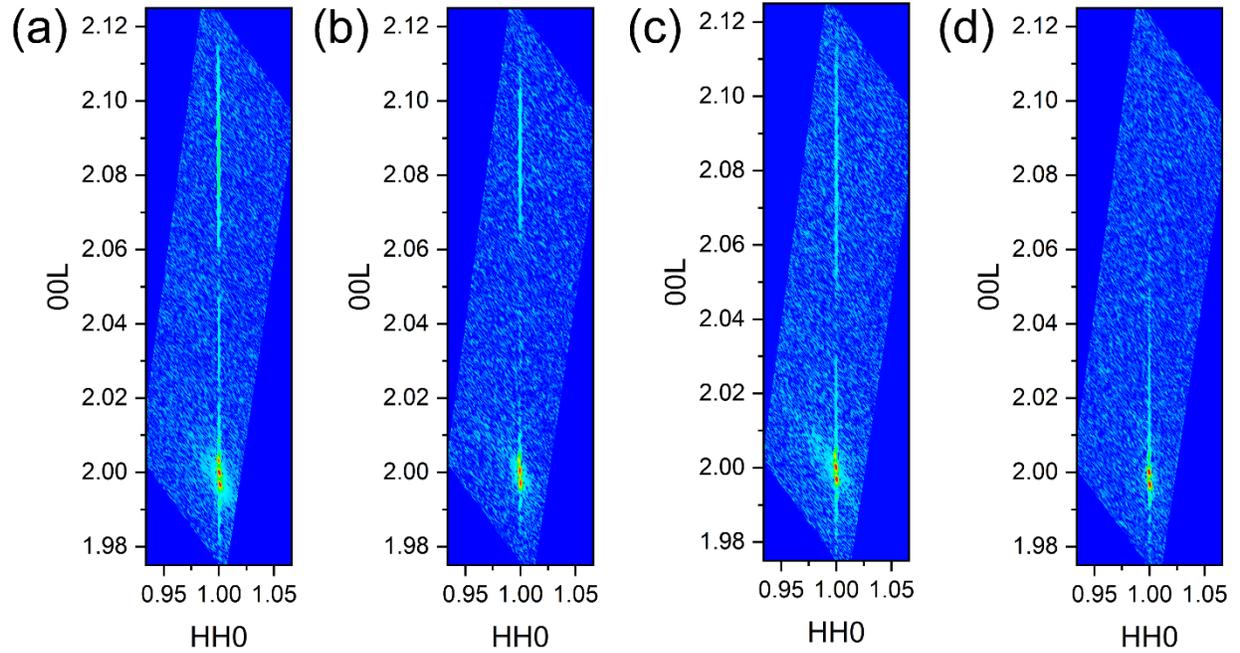

*Fig. S5. Asymmetrical RSM images around (112) plane for (**a**) 9.5 nm VO$_2$ thin film, and VO$_2$ heterostructures with VO$_2$ thicknesses of (**b**) 9.5 nm (**c**) 6.5 nm and (**d**) 1.5 nm. The vertical and horizontal axis of all the plots is scaled relative to the miller indices of the TiO$_2$ (001) substrate. The RSM data presented here is further evidence that there are no measurable changes to the unit cell volume across all heterostructures and that all films are coherently strained to the TiO$_2$ (001) substrate.*

## S6: Temperature-dependent sheet resistance of VO$_2$ thin films

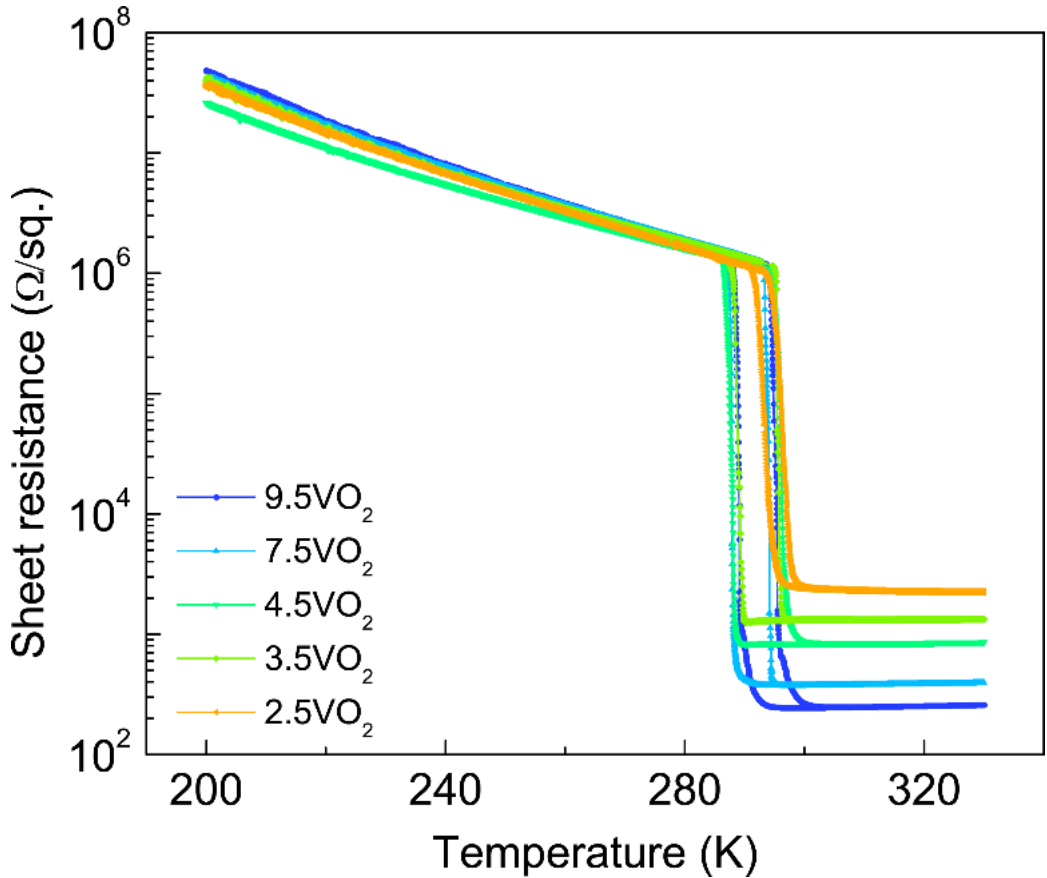

***Fig. S6.*** *Temperature-dependent sheet resistance of VO$_2$ thin films with varying VO$_2$ thicknesses as mentioned in the legends. For nomenclature simplicity, VO$_2$ thin films are written as tVO$_2$ where 't' is the thickness of VO$_2$. For example, 9.5VO$_2$ corresponds to a 9.5 nm thick VO$_2$ thin film. The increase in metallic state resistance as the thickness of VO$_2$ film is decreased is suggestive of interfacial scattering. This also shows all the films, except 2.5VO$_2$, have nearly the same transition temperature (~295 K) which again confirms the reduction in transition temperature in VO$_2$ heterostructures is entirely after the formation of the heterostructure. The slight increase in transition temperature for the 2.5VO$_2$ film is attributed to be due to titanium interdiffusion at the VO$_2$/TiO$_2$(substrate) interface.[4,5] We were not able to measure 1.5 nm VO$_2$ thin film as it was found to be not stable when exposed to ambient without a capping layer.*

# S7: Temperature-dependent sheet resistance of VO$_2$ thin film and heterostructures

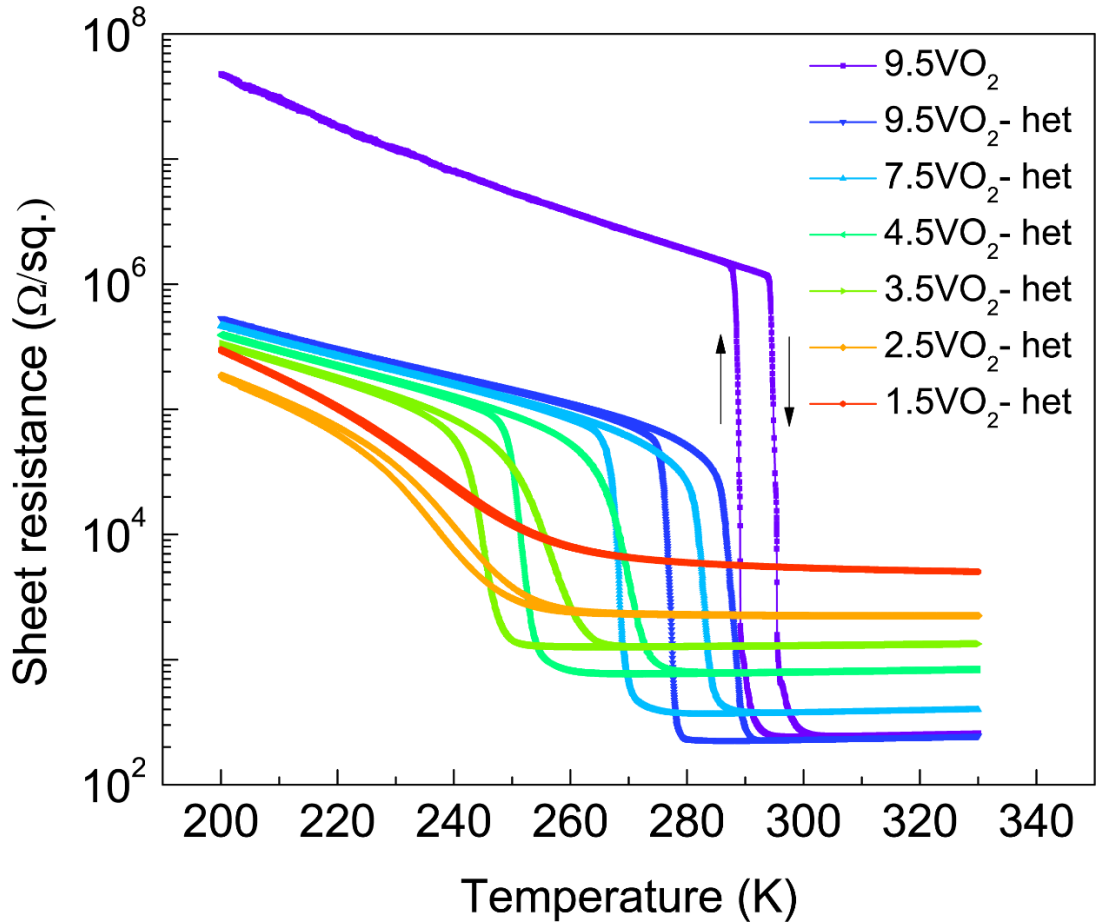

*Fig. S7. Temperature-dependent sheet resistance of 9.5 nm VO$_2$ thin film and VO$_2$ heterostructures with varying VO$_2$ thicknesses, t. For nomenclature simplicity, we distinguish VO$_2$ thin films and heterostructures with a VO$_2$ thickness of 't' as tVO$_2$ and tVO$_2$-het, respectively. For example, 9.5VO$_2$-het corresponds to a heterostructure with 9.5 nm thick VO$_2$. The sheet resistance data in this figure corresponds to the normalized resistance presented in Fig. 2b of the main text. The increase in metallic state resistance as the thickness of VO$_2$ film is decreased is suggestive of interfacial scattering. Importantly, it can be seen that the metallic state resistance is identical for the 9.5 nm VO$_2$ heterostructure and 9.5 nm VO$_2$ thin film. This is also an indirect indication of the sharp interfaces across the heterostructure layers; Rough interfaces could lead to increased interfacial scattering and a higher metallic state resistance for the 9.5 nm VO$_2$ heterostructure in comparison with the 9.5 nm VO$_2$ thin film.*

# S8: Calculation of $T_{MIT}$ from sheet resistance vs temperature curves

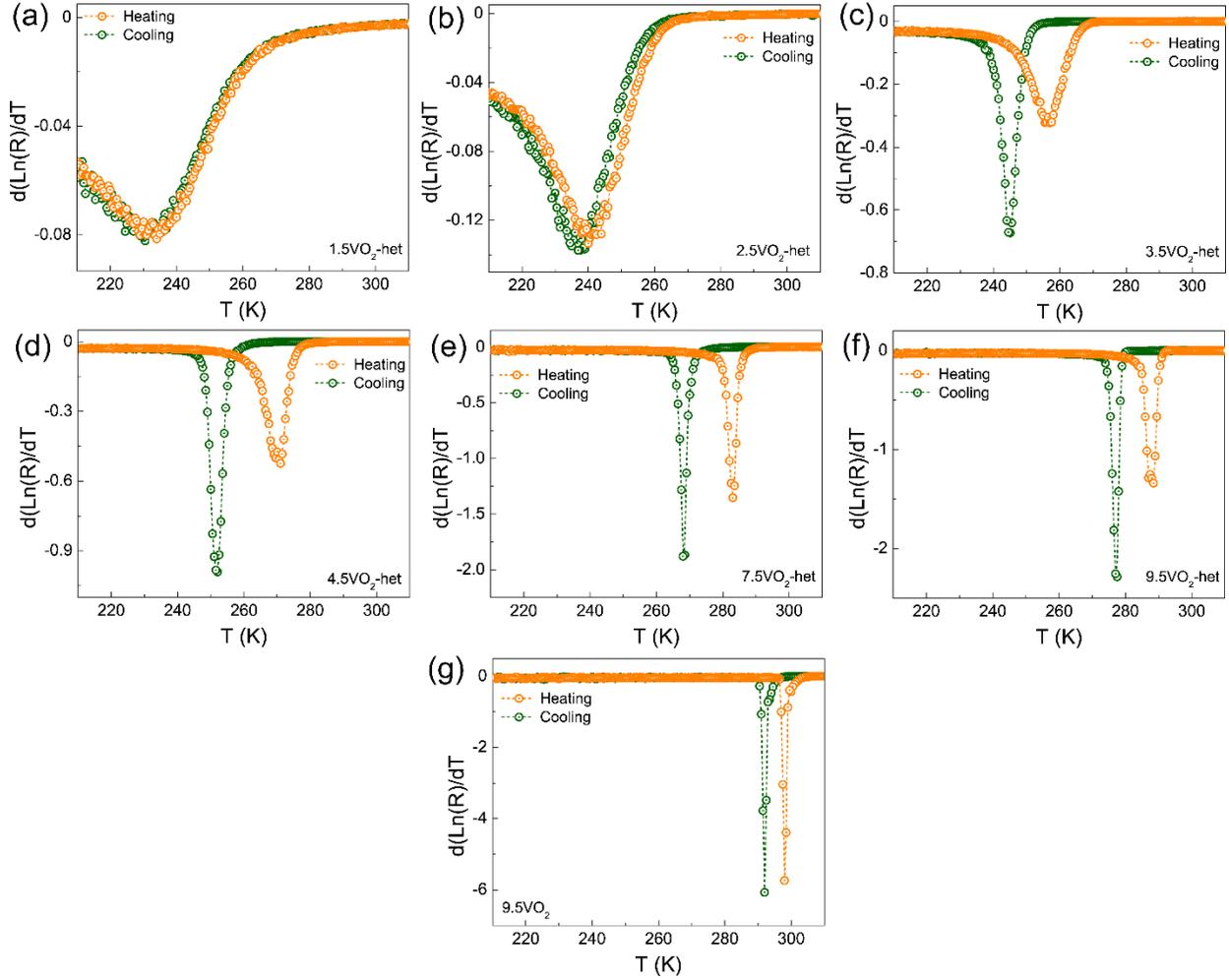

*Fig. S8. d(ln(R))/dT plots for both the heating and cooling cycles for VO₂ heterostructures with VO₂ thicknesses of (**a**) 1.5 nm, (**b**) 2.5 nm, (**c**) 3.5 nm, (**d**) 4.5 nm, (**e**) 7.5 nm, (**f**) 9.5 nm, and, (**g**) for a 9.5 nm VO₂ thin film. Here, ln(R) is the natural log of the temperature-dependent sheet resistance of the films as shown in Fig. S7 and d/dT is the derivative operator with respect to temperature, T. Peak positions of the d(ln(R))/dT plots correspond to the transition temperature for the respective thermal cycle. All the transition temperatures discussed in the main text are the average transition temperature of the transition temperature for heating and cooling cycle given by $T_{MIT} = ((T_{MIT})_{heating} + (T_{MIT})_{cooling})/2$. The estimated $T_{MIT}$ are: 233 K for 1.5 nm VO₂; 237 K for 2.5 nm VO₂; 250 K for 3.5 nm VO₂; 260 K for 4.5 nm VO₂; 275 K for 7.5 nm VO₂; 282 K for 9.5 nm VO₂ in VO₂ heterostrtuctures. By comparison, the $T_{MIT}$ is 295 K for 9.5 nm VO₂ thin film.*
9

## S9: Temperature-dependent sheet resistance of VO$_2$ heterostructures and controls

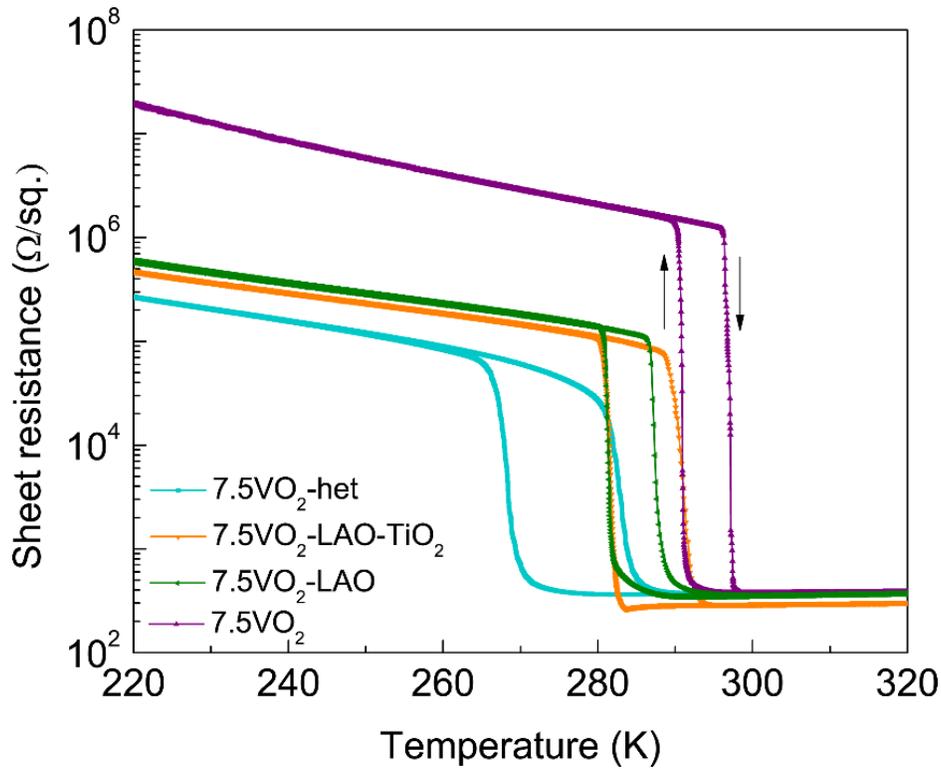

*Fig. S9.* A comparison of temperature-dependent sheet resistance characteristics of modulation-doped 7.5 nm VO$_2$ heterostructure (7.5VO$_2$-het) and 7.5 nm VO$_2$ thin film (7.5VO$_2$) with various VO$_2$ heterostructure 'controls'. The control heterostructures include 7.5 nm VO$_2$ thin film capped with 2 nm LAO (7.5VO$_2$-LAO) and a 7.5 nm VO$_2$ heterostructure with a near stoichiometric TiO$_2$ deposited at 10 mTorr of oxygen pressure with a 1 nm LAO capping layer over TiO$_2$ (7.5VO$_2$-LAO-TiO$_2$). The number prefixed with each of the layers represent the respective film thickness in nm. For the 7.5VO$_2$-LAO heterostructure, there is a reduction in resistivity in the insulating phase and a small but observable reduction in transition temperature ($T_{MIT}$) ~6 K compared to the $T_{MIT}$ of VO$_2$ thin film. However, for the modulation-doped heterostructure, (7.5VO$_2$-het), there is a further reduction in resistivity with a decrease in $T_{MIT}$ of ~20 K compared to the $T_{MIT}$ of VO$_2$ thin film. For the heterostructure where TiO$_2$ is near stoichiometric (7.5VO$_2$-LAO-TiO$_2$), the reduction in $T_{MIT}$ is also ~6 K, similar to the reduction in $T_{MIT}$ for 7.5VO$_2$-LAO. Clearly, a majority of the reduction in $T_{MIT}$ comes from the charge transfer from the TiO$_{2-x}$ dopant layer.

## S10: Carrier density and carrier mobility for 7.5 nm $VO_2$ thin films and heterostructures

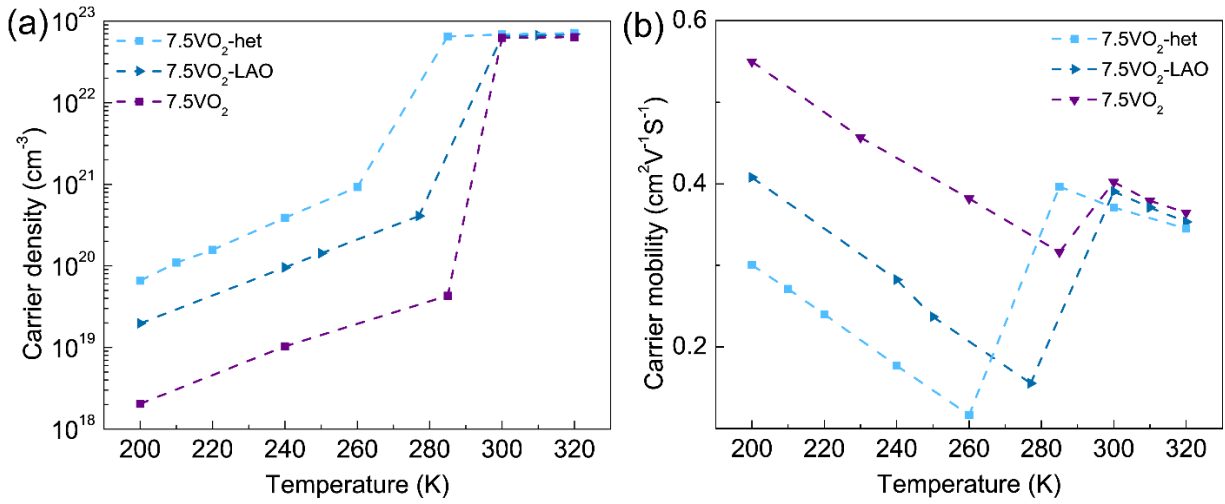

***Fig. S10.*** *Plots of temperature-dependent (**a**) carrier densities and (**b**) carrier mobilities for 7.5 nm $VO_2$ heterostructure (7.5$VO_2$-het), 7.5 nm $VO_2$ thin film (7.5$VO_2$) and 7.5 nm $VO_2$ capped with 2 nm LAO (7.5$VO_2$-LAO). For the 7.5$VO_2$-LAO, there is a noticeable increase in carrier density (and a decreased in carrier mobility) compared to 7.5 nm $VO_2$ thin film, but this is less than the carrier density observed in the comparable modulation-doped heterostructure (7.5$VO_2$-het).*

## S11: Binding energy calibration of $VO_2$ spectra across the MIT

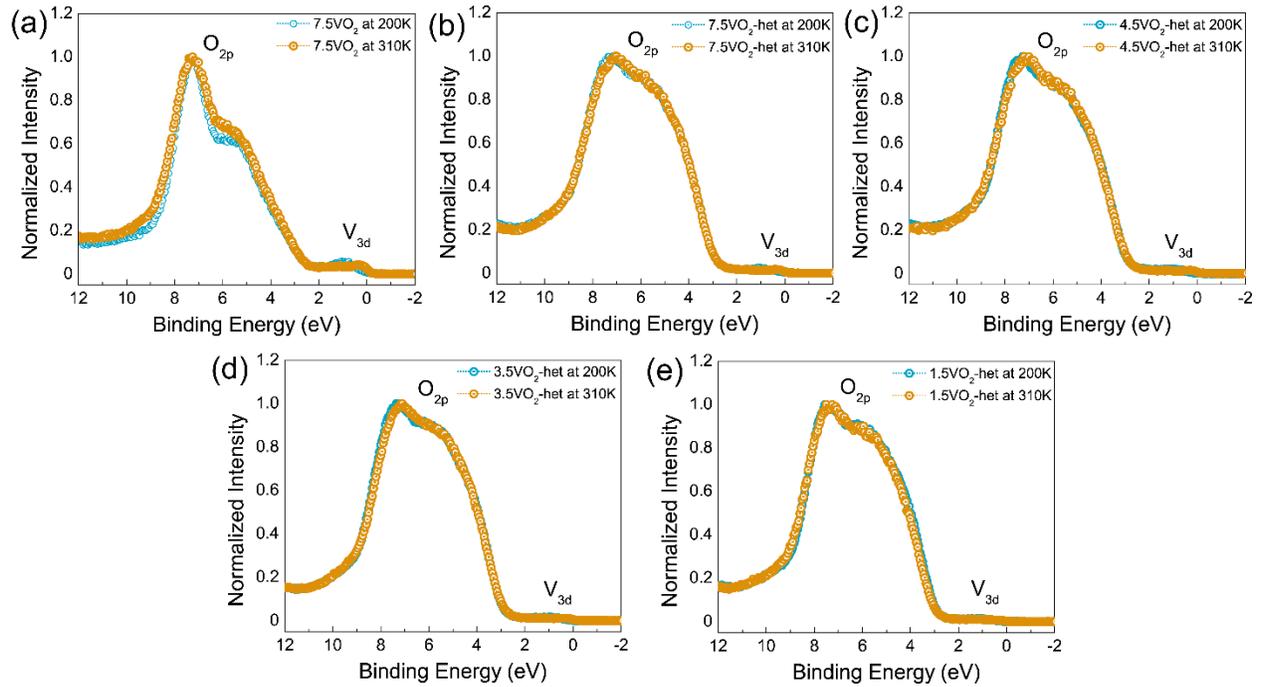

*Fig. S11. Binding energy calibration for the insulating state spectra. We noticed charging effects in some of the insulating state HAXPES spectra. In order to correct for charging effects, we used O 2p binding energy across insulating and metallic states of the samples as an internal reference. We note that metallic state spectra are unaffected by charging affects and several previous studies showed that binding energy of O 2p spectra does not change across the MIT[6–9]. Further, no changes to the O 2p contributions from the LAO and $TiO_2$ layers are expected across the MIT in $VO_2$. The valence band spectra for (**a**) 7.5 nm $VO_2$ film and $VO_2$ heterostructures with $VO_2$ thicknesses of (**b**) 7.5 nm, (**c**) 4.5 nm, (**d**) 3.5 nm, and (**e**) 1.5 nm are shown for both the metallic phase (at 310 K) and the insulating phase (at 200 K) after binding energy correction. All the V 2p spectra shown in the main text are based on this calibration.*

## S12: Summary of Binding energy changes for modulation-doped $VO_2$

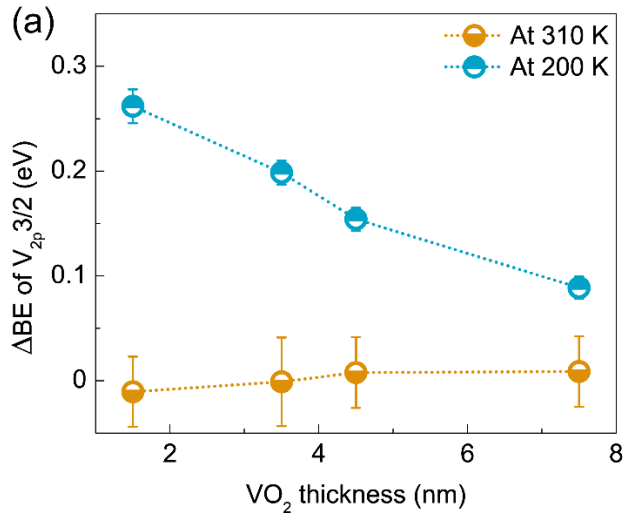

*Fig. S12.* *The change in binding energy of the V $2p_{3/2}$ peak for $VO_2$ heterostructures with respect to its binding energy in $VO_2$ thin films (ΔBE) for various thicknesses of $VO_2$ in heterostructures. There is no binding energy change in the metallic state (orange, at 310 K), while there is a binding energy increase for the same set of samples in the insulating state (blue, at 200 K). This is suggestive of band-bending leading to modulation-doping.*

## S13: Evolution of *P1* peak in metallic and insulating $VO_2$ heterostructures

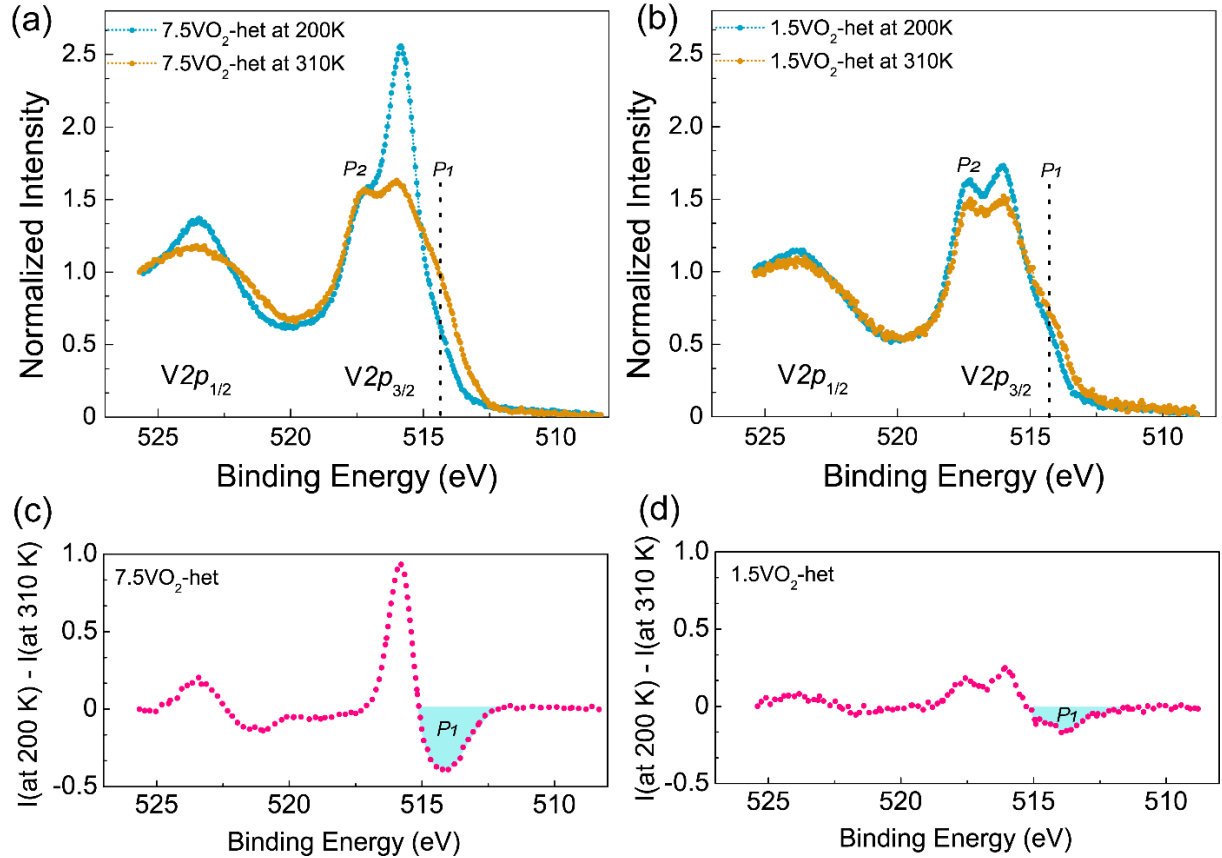

*Fig. S13.* V 2p core level spectra for (***a***) 7.5 nm and (***b***) 1.5 nm $VO_2$ heterostructures in the metallic (at 310 K) and insulating (at 200K) states normalized to the V 2p area under the curve. Corresponding intensity difference plots in (***c***) showing that the P1 peak (at ~514.5 eV) is prominent in the metallic phase for the 7.5 nm $VO_2$ heterostructure (negative intensity difference) while for the 1.5 nm heterostructure in (***d***) the intensity difference decreases suggesting an increase in the P1 intensity for the insulating state spectrum for the 1.5 nm $VO_2$ heterostructure.

## S14: HAXPES spectra of V $2p_{3/2}$

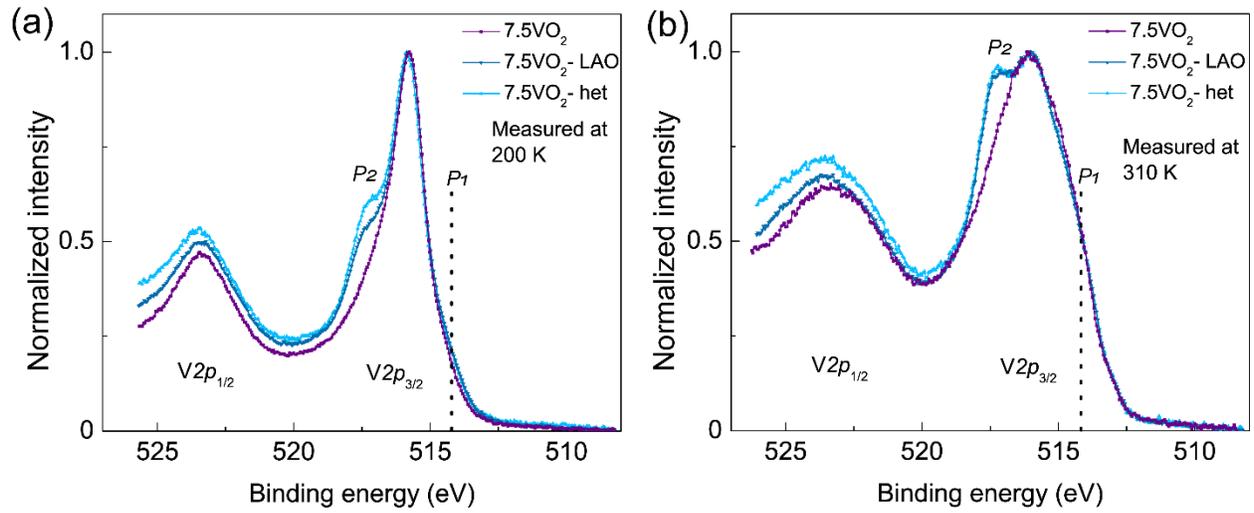

*Fig. S14.* A comparison of V 2p core-level spectra of 7.5 nm modulation doped $VO_2$ heterostructure, 7.5 nm $VO_2$ with 2 nm LAO capping layer and 7.5 nm $VO_2$ thin film in (*a*) the insulating (200 K) and (*b*) the metallic states (310 K). The additional peak P2 is observed after the deposition of LAO. However, the change in the peak P1 in the insulating state spectra is not significant.

## S15: HAXPES spectra of La *3d* and Al *1s*

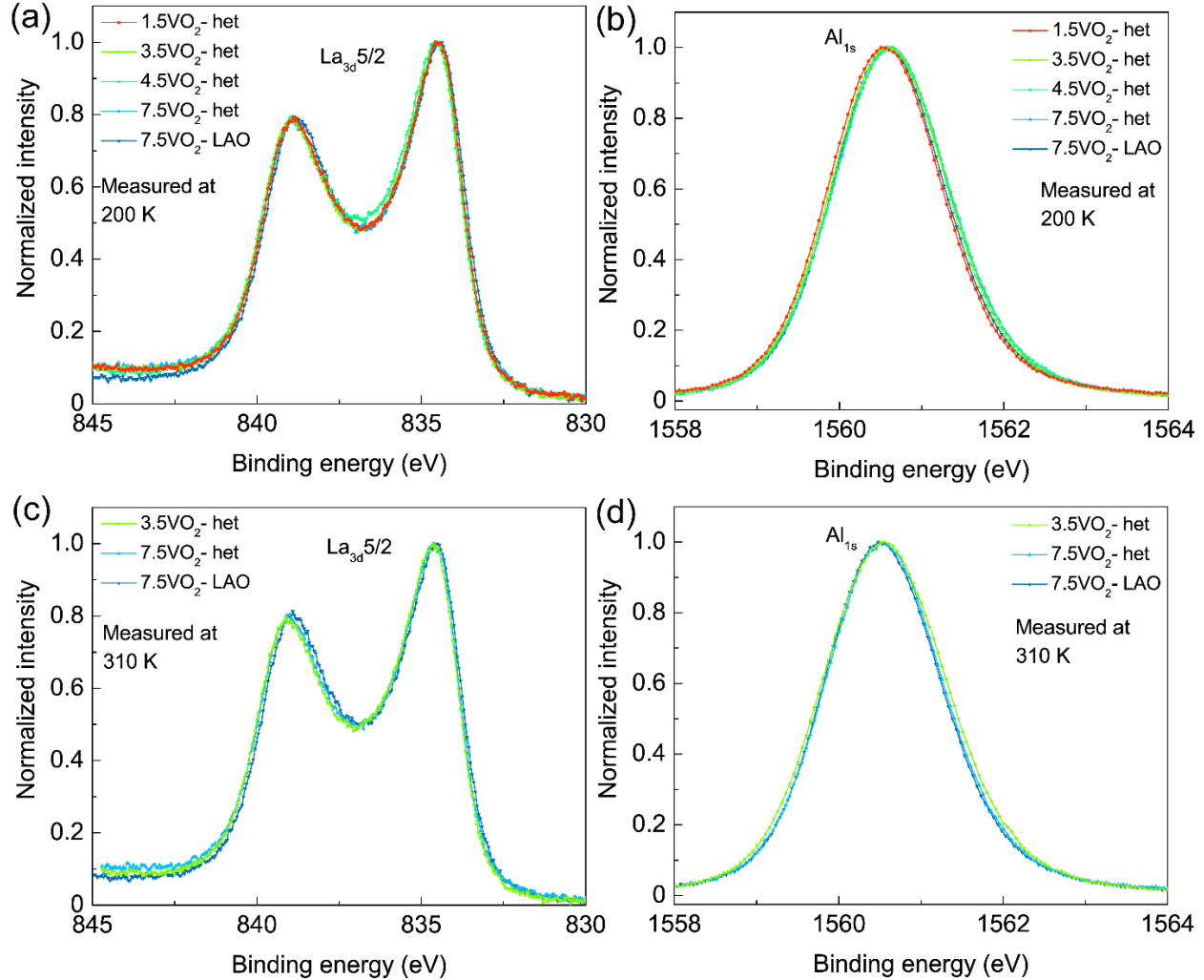

*Fig. S15.* *A comparison of core level HAXPES spectra of modulation doped $VO_2$ heterostructures of (**a**) La $3d_{5/2}$ and (**b**) Al 1s at 200 K, and (**c**) La $3d_{5/2}$ (**d**) Al 1s at 310 K. In the insulating state (at 200 K), the spectra were collected for $VO_2$ heterostructures corresponding to all $VO_2$ thicknesses used in this study. In the metallic state (at 310 K), the spectra were collected for 3.5 nm and 7.5 nm $VO_2$ heterostructures and for $7.5VO_2/2LAO$ heterostructure. No significant changes to the La and Al core level were observed.*

# S16: Valence band spectra in insulating state for VO₂ heterostructures and thin films

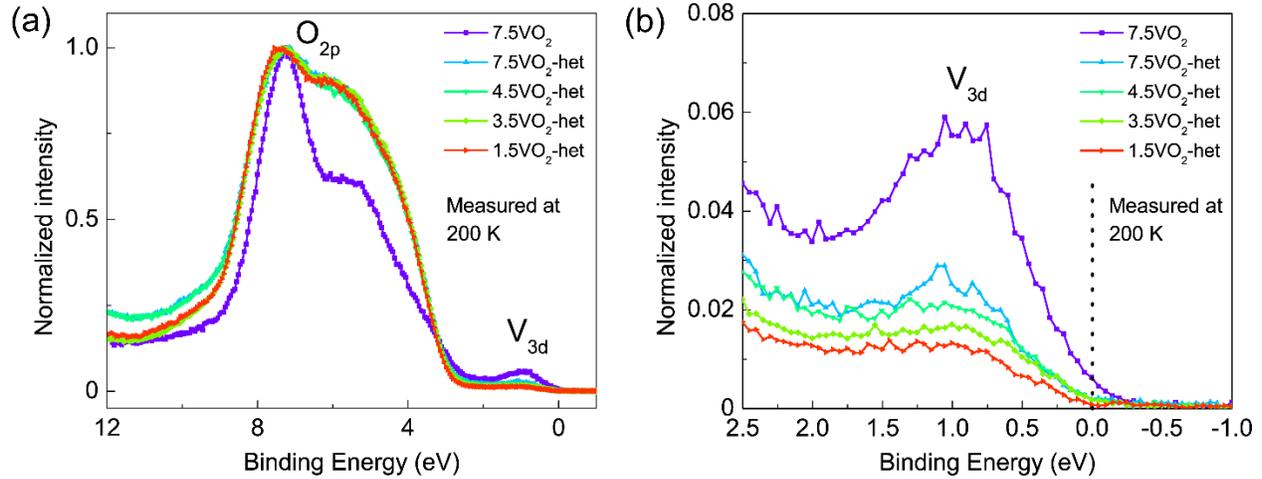

***Fig. S16.*** *(**a**) Valence band spectra (O 2p and V 3d) and, (**b**) V 3d spectra (zoom in of figure S16 a) measured at 200 K are shown here after the binding energy correction as discussed in S11. The V 3d peak position at ~0.9 eV is in good agreement with existing literature.*[6–9] *The vertical dotted line at 0 eV corresponds to the Fermi level.*

## Supplementary References